\documentclass[letterpaper,twocolumn,10pt]{article}
\usepackage{usenix-2020-09}

\usepackage[english]{babel}
\usepackage[T1]{fontenc}
\usepackage[utf8]{inputenc}

\usepackage{tikz}
\usepackage{csquotes}
\usepackage{cleveref}
\usepackage{siunitx}
\usepackage{listings}
\usepackage{subcaption}
\usepackage{booktabs}
\usepackage{adjustbox}
\usepackage[shortcuts]{extdash}
\usepackage{subcaption}	

\usepackage{tikz}
\usepackage{tikzsymbols}
\usetikzlibrary{calc,
	positioning,
	arrows.meta,
	shapes.geometric,
	decorations,
	patterns,
	arrows,
	  backgrounds,
	  fit,
	  patterns.meta,
	  chains,
	  scopes,
	  matrix,
	  arrows.meta,
}
\usetikzlibrary{shapes,arrows}
\usetikzlibrary{decorations.pathreplacing}
\usetikzlibrary{babel}
\usetikzlibrary{calc,arrows.meta,patterns,backgrounds}
\usetikzlibrary{shapes.multipart}

\begin{document}

\date{}

\title{CoRD: Converged RDMA Dataplane for High-Performance Clouds}

\author{
	{\rm Maksym Planeta}\\
	Barkhausen Institut
	\and
	{\rm Jan Bierbaum}\\
	TU Dresden
	\and
	{\rm Michael Roitzsch}\\
	Barkhausen Institut
	\and
	{\rm Hermann Härtig}\\
	TU Dresden
} 

\maketitle

\begin{abstract}
High-performance networking is often characterized by kernel bypass which is
considered mandatory in high-performance parallel and distributed applications.
But kernel bypass comes at a price because it breaks the traditional OS
architecture, requiring applications to use special APIs and limiting the OS
control over existing network connections.
We make the case, that kernel bypass is \emph{not} mandatory.
Rather, high-performance networking relies on multiple performance-improving
techniques, with kernel bypass being the least effective.
CoRD removes kernel bypass from RDMA networks, enabling efficient OS-level
control over RDMA dataplane.
\end{abstract}

\section{Introduction}
\label{sec:intro}

High Performance Computing (HPC) and Cloud architectures are converging: the
former is advancing towards greater elasticity and higher resource
utilization~\cite{elmaghraouiDynamicMalleabilityIterative2007,
chadhaExtendingSLURMDynamic2020, frankEffectsBenefitsNode2019, goldrush}, while
the latter is increasingly geared towards handling high\-/performance
workloads~\cite{carreiraCirrusServerlessFramework2019,
fengExploringServerlessComputing2018, puShufflingFastSlow2019}.
However, current HPC Cloud systems remain partitioned: vast sections lack
high-performance networking support, with only a few segments offering such
capabilities~\cite{HighPerformanceComputing2020, ElasticFabricAdapter,
oracleComputeShapes2023}.
As we push for a holistic integration of HPC and Cloud systems, we realize that
their network architectures must converge as well.
In this paper, we show how operating systems (OS) can help in such integration
by making high-performance networks more accessible.
We believe this approach can significantly enhance performance within Cloud
environments.

A classical OS provides a socket-based abstraction layer between the
applications and the underlying hardware.
This layer allows \enquote{rich and robust} OS services~\cite[p.\
228]{abranchesGettingBackWhat2022}, like application portability,
resource scheduling, and security policy enforcement.
High-performance networks, often represented by Remote Direct Memory Access
(RDMA) networks~\cite{dpdk, ibta} subvert this layered OS architecture by
granting user-level applications direct access to network devices (NICs).
These applications circumvent OS services (including the network stack), instead
they come with their own NIC drivers and perform most of the scheduling and
resource management themselves~\cite{dpdk, openmpi-docs}.
This approach forces the OS to rely on the NIC to implement security
(IOMMUs~\cite{redmark, srdma} or VLANs~\cite{simpsonSecuringRDMAHighPerformance}
are not sufficient) and resource-sharing policies~\cite{justitia,freeflow,masq}.
As a result, the OS becomes just an over-engineered bootloader~\cite{ihk, mos,
cnk, decoupled}.

In response to the demand for flexibility and OS control over high-performance
network communication, the OS community has proposed several \emph{dataplane}
architectures~\cite{ix, demikernel, dpdk, masq, snap,
pfefferleHybridVirtualizationFramework2015, iouring, tsaiLITEKernelRDMA2017,
mouzakitisLightweightGenericRDMA2017}.
But few of these solutions have seen widespread adoption~\cite{dpdk,
iouring}.
The novelty of these architectures and their lack of interoperability with
socket-based communication, poses a significant entry barrier, restricting their
appeal for conventional Cloud applications.
Interestingly, dataplane architectures and traditional RDMA networks share
common mechanisms and principles, suggesting a potential for compatibility.
So, could we introduce minor, strategic adjustments to the widely-used RDMA
networks to make them as flexible and controllable by the OS as socket-based
networks?
This approach could bridge the current gap and facilitate the convergence of HPC
and Cloud architectures.

%

In this paper, we identify a set of key properties that boost the performance of
RDMA networks over socket-based networks.
Among these, kernel-bypass is often considered the cornerstone~\cite{anycall,
elphinstoneL3SeL4What2013, singularity, unikernels}.
However, we challenge this prevalent belief and show that kernel-bypass is
not indispensible to high-performance applications.
Inspired by these insights, we propose the Converged RDMA Dataplane (CoRD), a
strategically altered version of the traditional RDMA architecture that directs
the dataplane through the kernel.
This pivotal shift gives the kernel full control over RDMA communication,
enabling it to enforce security policies~\cite{apparmor, selinux}, provide
virtualization~\cite{mianoFrameworkEBPFBasedNetwork2021, masq}, control
resources on a fine-grained level~\cite{freeflow, masq}, and facilitate
application observability~\cite{abranchesEfficientNetworkMonitoring2021}.
Preliminary measurements reveal that CoRD retains the performance for
large-scale applications, effectively converging RDMA and socket-based networks
without forsaking the intrinsic advantages of either.


\section{RDMA performance breakdown}
\label{sec:motivation}

Compared to traditional socket-based networks, we identify three software
properties enabling high\-/performance communication over RDMA networks:
\emph{kernel-bypass} (or \emph{OS-bypass}), \emph{zero-copy}, and
\emph{polling}\footnote{There are also hardware techniques, like \emph{one-sided
(or RDMA) communication}, \emph{offloading}, and \emph{lossless congestion
control}.}.
Each technique improves performance in a different way, can be employed
independently of the rest, and comes with its own limitations.
This diversity suggests that these techniques may not be equally performance
critical in every use case.

\emph{Kernel-bypass} allows the application to access the device directly,
avoiding overhead for crossing the user-kernel boundary.
Zero-copy enables the NIC to access the application memory directly, avoiding
overhead for copying the message content between kernel and user memory as done
with \texttt{read} and \texttt{write} system calls.
With polling, the application continuously checks the NIC's message queues to
avoid the overhead of interrupt processing, which happens in the \texttt{epoll}
system call.
Together, these techniques form an API~\cite{ibverbs,demikernel,ugni} that is
very different from the traditional socket-based API.

A simple example illustrates the role of each of the aforementioned techniques:
A high-performance \texttt{perftest} microbenchmark measures (on our local
system \emph{L}, see~\cref{sec:eval}) point-to-point send latency and throughput
for different message sizes.
To quantify the contribution of each technique, we modified the benchmark to
\enquote{remove} individual techniques from the typical RDMA mode of operation
and measure how the benchmark performance changes.
To emulate the removal of zero-copy, our \texttt{perftest} makes an extra memory
copy when sending or receiving a message.
We emulate the removal of kernel-bypass by adding a \texttt{getppid} system
call.
Finally, we instruct the benchmarks to wait for an interrupt from the NIC
instead of busy-polling, an existing, but rarely used feature in RDMA networks.
%

\begin{figure}
    \centering
    \begin{subcaptionblock}{.39\linewidth}
        \centering
        \begin{tabular}{ccc}\toprule
            \small $\SI{16}{\byte}$ & \small $\SI{4}{\kibi\byte}$ & \small $\SI{1}{\mebi\byte}$ \\\midrule
            \multicolumn{3}{l}{\small Baseline}                                                 \\
            $0.99$                  & $1.95$                      & $86$                        \\\cmidrule(lr){1-3}
            \multicolumn{3}{l}{\small No kernel bypass}                                         \\
            $1.06$                  & $1.95$                      & $86$                        \\\cmidrule(lr){1-3}
            \multicolumn{3}{l}{\small No busy-polling}                                          \\
            $4.69$                  & $4.16$                      & $90$                        \\\cmidrule(lr){1-3}
            \multicolumn{3}{l}{\small No zero copy (ZC)}                                        \\
            $1.03$                  & $2.31$                      & $229$                       \\
            \bottomrule
        \end{tabular}
        \caption{Latency (\unit{\micro\second})}\label{tab:remove-lat}
    \end{subcaptionblock}%
    \hfill%
    \begin{subcaptionblock}{.57\linewidth}
        \centering
\begin{tikzpicture}[x=1pt,y=1pt]
\definecolor{fillColor}{RGB}{255,255,255}
\path[use as bounding box,fill=fillColor,fill opacity=0.00] (0,0) rectangle (122.86,155.38);
\begin{scope}
\path[clip] (  0.00,  0.00) rectangle (122.86,155.38);
\definecolor{drawColor}{RGB}{255,255,255}
\definecolor{fillColor}{RGB}{255,255,255}

\path[draw=drawColor,line width= 0.5pt,line join=round,line cap=round,fill=fillColor] (  0.00,  0.00) rectangle (122.86,155.38);
\end{scope}
\begin{scope}
\path[clip] ( 33.61, 25.90) rectangle (117.86,150.38);
\definecolor{fillColor}{RGB}{255,255,255}

\path[fill=fillColor] ( 33.61, 25.90) rectangle (117.86,150.38);
\definecolor{fillColor}{RGB}{253,192,134}

\path[fill=fillColor] ( 62.33, 31.55) rectangle ( 73.82,104.32);
\definecolor{fillColor}{RGB}{190,174,212}

\path[fill=fillColor] ( 50.84, 31.55) rectangle ( 62.33,117.37);
\definecolor{fillColor}{RGB}{127,201,127}

\path[fill=fillColor] ( 39.35, 31.55) rectangle ( 50.84,111.97);
\definecolor{fillColor}{RGB}{253,192,134}

\path[fill=fillColor] (100.63, 31.55) rectangle (112.11,131.69);
\definecolor{fillColor}{RGB}{190,174,212}

\path[fill=fillColor] ( 89.14, 31.55) rectangle (100.63,129.45);
\definecolor{fillColor}{RGB}{127,201,127}

\path[fill=fillColor] ( 77.65, 31.55) rectangle ( 89.14, 71.07);
\definecolor{drawColor}{RGB}{0,0,0}

\node[text=drawColor,rotate= 90.00,anchor=base west,inner sep=0pt, outer sep=0pt, scale=  0.77] at ( 70.72, 35.56) {No kernel bypass};

\node[text=drawColor,rotate= 90.00,anchor=base west,inner sep=0pt, outer sep=0pt, scale=  0.77] at ( 59.23, 35.56) {No busy-polling};

\node[text=drawColor,rotate= 90.00,anchor=base west,inner sep=0pt, outer sep=0pt, scale=  0.77] at ( 47.74, 35.56) {No ZC};

\node[text=drawColor,rotate= 90.00,anchor=base west,inner sep=0pt, outer sep=0pt, scale=  0.77] at (109.02, 35.56) {No kernel bypass};

\node[text=drawColor,rotate= 90.00,anchor=base west,inner sep=0pt, outer sep=0pt, scale=  0.77] at ( 97.53, 35.56) {No busy-polling};

\node[text=drawColor,rotate= 90.00,anchor=base west,inner sep=0pt, outer sep=0pt, scale=  0.77] at ( 86.04, 35.56) {No ZC};
\definecolor{drawColor}{RGB}{0,0,0}

\path[draw=drawColor,draw opacity=0.60,line width= 0.6pt,dash pattern=on 4pt off 4pt ,line join=round] ( 33.61,131.70) -- (117.86,131.70);

\node[text=drawColor,text opacity=0.60,anchor=base,inner sep=0pt, outer sep=0pt, scale=  0.85] at ( 56.58,133.77) {1.4 Gb/s};

\node[text=drawColor,text opacity=0.60,anchor=base,inner sep=0pt, outer sep=0pt, scale=  0.85] at ( 94.88,133.77) {100 Gb/s};

\node[text=drawColor,text opacity=0.60,anchor=base,inner sep=0pt, outer sep=0pt, scale=  0.85] at ( 75.73,141.78) {Baseline BW};
\end{scope}
\begin{scope}
\path[clip] (  0.00,  0.00) rectangle (122.86,155.38);
\definecolor{drawColor}{RGB}{0,0,0}

\path[draw=drawColor,line width= 0.6pt,line join=round] ( 33.61, 25.90) --
	( 33.61,150.38);
\end{scope}
\begin{scope}
\path[clip] (  0.00,  0.00) rectangle (122.86,155.38);
\definecolor{drawColor}{RGB}{0,0,0}

\node[text=drawColor,anchor=base east,inner sep=0pt, outer sep=0pt, scale=  1.00] at ( 29.11, 28.11) {0.0};

\node[text=drawColor,anchor=base east,inner sep=0pt, outer sep=0pt, scale=  1.00] at ( 29.11, 78.19) {0.5};

\node[text=drawColor,anchor=base east,inner sep=0pt, outer sep=0pt, scale=  1.00] at ( 29.11,128.26) {1.0};
\end{scope}
\begin{scope}
\path[clip] (  0.00,  0.00) rectangle (122.86,155.38);
\definecolor{drawColor}{gray}{0.20}

\path[draw=drawColor,line width= 0.5pt,line join=round] ( 31.11, 31.55) --
	( 33.61, 31.55);

\path[draw=drawColor,line width= 0.5pt,line join=round] ( 31.11, 81.63) --
	( 33.61, 81.63);

\path[draw=drawColor,line width= 0.5pt,line join=round] ( 31.11,131.70) --
	( 33.61,131.70);
\end{scope}
\begin{scope}
\path[clip] (  0.00,  0.00) rectangle (122.86,155.38);
\definecolor{drawColor}{RGB}{0,0,0}

\path[draw=drawColor,line width= 0.6pt,line join=round] ( 33.61, 25.90) --
	(117.86, 25.90);
\end{scope}
\begin{scope}
\path[clip] (  0.00,  0.00) rectangle (122.86,155.38);
\definecolor{drawColor}{gray}{0.20}

\path[draw=drawColor,line width= 0.5pt,line join=round] ( 56.58, 23.40) --
	( 56.58, 25.90);

\path[draw=drawColor,line width= 0.5pt,line join=round] ( 94.88, 23.40) --
	( 94.88, 25.90);
\end{scope}
\begin{scope}
\path[clip] (  0.00,  0.00) rectangle (122.86,155.38);
\definecolor{drawColor}{RGB}{0,0,0}

\node[text=drawColor,anchor=base,inner sep=0pt, outer sep=0pt, scale=  0.80] at ( 56.58, 17.11) {16 B};

\node[text=drawColor,anchor=base,inner sep=0pt, outer sep=0pt, scale=  0.80] at ( 94.88, 17.11) {16 MiB};
\end{scope}
\begin{scope}
\path[clip] (  0.00,  0.00) rectangle (122.86,155.38);
\definecolor{drawColor}{RGB}{0,0,0}

\node[text=drawColor,anchor=base,inner sep=0pt, outer sep=0pt, scale=  1.00] at ( 75.73,  6.94) {Message Size, b};
\end{scope}
\begin{scope}
\path[clip] (  0.00,  0.00) rectangle (122.86,155.38);
\definecolor{drawColor}{RGB}{0,0,0}

\node[text=drawColor,rotate= 90.00,anchor=base,inner sep=0pt, outer sep=0pt, scale=  1.00] at ( 11.89, 88.14) {Relative Bandwidth};
\end{scope}
\end{tikzpicture}
        \vspace{-1em}
        \caption{Bandwidth}\label{fig:remove-bw}
    \end{subcaptionblock}%
    \caption{%
        \enquote{Removing} performance-improving techniques compared to having all techniques active
    }%
    \label{fig:remove}
\end{figure}

\Cref{fig:remove} shows the results of our experiment.
Removing any of the three techniques has a significant influence on
small-message throughput (\cref{fig:remove-bw}), because this task is CPU-bound.
Even the baseline variant (with all techniques active) achieves only
$\SI{1.4}{\giga\bit/\second}$ out of the theoretical maximum of
$\SI{100}{\giga\bit/\second}$.
On the other hand, large message throughput is obstructed significantly only
without zero-copy.
For latency (\cref{tab:remove-lat}), removing busy-polling adds significant
overhead from additional interrupt processing, especially for small messages,
although the absolute overhead stays the same even for very large messages.
Removing zero-copy adds additional latency proportional to message size (up to
$\SI{140}{\micro\second/\mebi\byte}$) because of additional data movement.
Finally, removing kernel bypass adds small constant overhead with minimal
overall impact.

We make the following observations from this experiment:
Zero copy appears to be the largest contributor to performance, especially for
large messages.
Kernel-bypass is crucial for throughput and latency only for very small
messages.
Even then, polling is more important than kernel-bypass.
These observations identify kernel-bypass as the least performance-critical
technique.

\section{Converged RDMA Dataplane}
\label{sec:model}

\newcommand{\pict}[2]{\tikz[baseline=-2.5pt]{\node[#1,minimum width=0.7em, inner sep = .75pt,
      minimum height=0.1em, text height=1.2ex, font=\footnotesize]{#2};}}

\newcommand{\basewidth}{1.1cm}
\newcommand{\swheight}{0.98cm}
\newcommand{\baseskip}{3pt}

\definecolor{usermem}{RGB}{239,237,245}
\definecolor{nicmem}{RGB}{141,160,203}
\definecolor{control}{RGB}{61,138,185}
\definecolor{data}{RGB}{255,80,80}
\definecolor{ulevel}{RGB}{182,205,157}
\definecolor{dlevel}{RGB}{161,215,106}

\tikzset{
sw/.style = {
    node distance = 0pt,
  },
user/.style = {
    top color=ulevel!60,
    bottom color=ulevel,
    rectangle,shade,
    rounded corners,
  },
kernel/.style = {
user,
minimum width = (\basewidth - 0.3cm),
minimum height = \swheight,
node distance = \baseskip,
label={[anchor=base]above,label distance=0.7ex:{\emph{\small Kernel}}},
},
app/.style = {
user,
minimum width = (\basewidth + 0.3cm),
minimum height = \swheight,
node distance = \baseskip,
label={[anchor=base]above,label distance=0.7ex:{\emph{\small Application}}},
},
mem/.style = {
    minimum height=10pt,
    minimum width=10pt,
    node distance = 0pt,
    rounded corners=1pt,
  },
umem/.style = {
    mem,
    fill=usermem,
  },
kmem/.style = {
    mem,
    fill=nicmem,
  },
kumem/.style = {
    umem,
    fill=nicmem,
  },
nic/.style = {
    user,
    top color=dlevel!40,
    bottom color=dlevel,
    node distance = 12pt,
    minimum width = (\basewidth + \basewidth + \baseskip + \baseskip),
    minimum height = 2.5ex
  },
flow/.style = {
rounded corners, line width = 1pt, color=data, -{Latex}
},
control/.style = {
rounded corners, line width = 0.8pt, color=control, -{Latex}
},
control point/.style = {
    draw,
    minimum size=3pt,
    align=center,
    inner sep=0pt,
    fill=black,
    node distance = 4pt,
    circle,
  },
flow step/.style= {
    draw,
    circle ,
    inner sep=0.1pt,
    font=\footnotesize,
  },
control step/.style={
    flow step,
    fill=control,
    color=control,
    text=black,
  },
data step/.style={
    flow step,
    fill = data,
    color = data,
    text = black,
  }
}

\tikzset{
  diagonal fill/.style 2 args={fill=#2, path picture={
          \fill[#1, sharp corners] (path picture bounding box.south west) -|
          (path picture bounding box.north east) -- cycle;}},
  reversed diagonal fill/.style 2 args={fill=#2, path picture={
          \fill[#1, sharp corners] (path picture bounding box.north west) |-
          (path picture bounding box.south east) -- cycle;}}
}

\begin{figure}
    \centering
    \begin{subfigure}[t]{0.3\linewidth}
        \tikzset{
sw/.style = {
    node distance = 0pt,
  },
user/.style = {
    top color=ulevel!60,
    bottom color=ulevel,
    rectangle,shade,
    rounded corners,
  },
kernel/.style = {
user,
minimum width = (\basewidth),
minimum height = \swheight,
node distance = \baseskip,
label={[anchor=base]above,label distance=0.7ex:{\emph{\small Kernel}}},
},
app/.style = {
user,
minimum width = (\basewidth),
minimum height = \swheight,
node distance = \baseskip,
label={[anchor=base]above,label distance=0.7ex:{\emph{\small User}}},
},
mem/.style = {
    minimum height=10pt,
    minimum width=10pt,
    node distance = 0pt,
    rounded corners=1pt,
  },
kmem/.style = {
    mem,
    fill=nicmem,
  },
kumem/.style = {
    umem,
    fill=nicmem,
  },
nic/.style = {
    user,
    top color=dlevel!40,
    bottom color=dlevel,
    node distance = 12pt,
    minimum width = (\basewidth + \basewidth + \baseskip + \baseskip),
    minimum height = 2.5ex
  },
flow/.style = {
rounded corners, line width = 1pt, color=data, -{Latex}
},
control/.style = {
rounded corners, line width = 0.8pt, color=control, -{Latex}
},
control point/.style = {
    draw,
    minimum size=3pt,
    align=center,
    inner sep=0pt,
    color=black,
    fill=black,
    node distance = 4pt,
    circle,
  },
flow step/.style= {
    draw,
    circle ,
    text=black,
    fill=control,
    inner sep=0.1pt,
    font=\footnotesize,
  },
control step/.style={
    flow step,
    text=black,
    fill=control,
  },
data step/.style={
    flow step,
    text = black,
    fill = data,
  }
}

\begin{tikzpicture}
  \node[sw] (sw1) {};
  \node[kernel] (k1) [right = of sw1.center] {};
  \node[app] (a1) [left = of sw1.center] {};

  \node[umem] (amem1) at ($(a1.east) + (-0.8cm,-0.1cm)$) {};
  \node[kmem] (kmem1) at ($(k1.west) + (.5cm,-0.1cm)$) {};

  \node[nic, label={[anchor = west]left:{\small NIC}}] (n1) [below = of a1.south west, anchor=west] {};
  \node[sw, node distance=(\basewidth + 0.6cm)] (sw2) [right = of k1] {};
  \coordinate (i1) at ($(kmem1 |- n1.south) -(0,0.3cm) $);

  \node[control point] (c1) [above right = of amem1] {};

  \node[control point] (c2) [above right = of kmem1] {};
  \node[control point] (c2n) at (n1-|c2) {};
  \coordinate (c3) at ($(c2 |- i1) -  (0,0.2cm)$);

  \coordinate (c4) at (c3|-i1);

  \draw[flow] (amem1) to node[data step] {2} (kmem1);
  \draw[flow] (kmem1)  to node[data step] {4} (i1);

  \draw[control] (c1) --  node[control step] {1} (c2);
  \draw[control] (c2)  --   node[control step] {3} (c2n);


\end{tikzpicture}%
        \caption{Sockets}%
        \label{fig:network-traditional}
    \end{subfigure}
    ~
    \begin{subfigure}[t]{0.3\linewidth}
        \tikzset{
sw/.style = {
    node distance = 0pt,
  },
user/.style = {
    top color=ulevel!60,
    bottom color=ulevel,
    rectangle,shade,
    rounded corners,
  },
kernel/.style = {
user,
minimum width = (\basewidth),
minimum height = \swheight,
node distance = \baseskip,
label={[anchor=base]above,label distance=0.7ex:{\emph{\small Kernel}}},
},
app/.style = {
user,
minimum width = (\basewidth),
minimum height = \swheight,
node distance = \baseskip,
label={[anchor=base]above,label distance=0.7ex:{\emph{\small User}}},
},
mem/.style = {
    minimum height=10pt,
    minimum width=10pt,
    node distance = 0pt,
    rounded corners=1pt,
  },
kmem/.style = {
    mem,
    fill=nicmem,
  },
kumem/.style = {
    umem,
    fill=nicmem,
  },
nic/.style = {
    user,
    top color=dlevel!40,
    bottom color=dlevel,
    node distance = 12pt,
    minimum width = (\basewidth + \basewidth + \baseskip + \baseskip),
    minimum height = 2.5ex
  },
flow/.style = {
rounded corners, line width = 1pt, color=data, -{Latex}
},
control/.style = {
rounded corners, line width = 0.8pt, color=control, -{Latex}
},
control point/.style = {
    draw,
    minimum size=3pt,
    align=center,
    inner sep=0pt,
    color=black,
    fill=black,
    node distance = 4pt,
    circle,
  },
flow step/.style= {
    draw,
    circle ,
    text=black,
    fill=control,
    inner sep=0.1pt,
    font=\footnotesize,
  },
control step/.style={
    flow step,
    text=black,
    fill=control,
  },
data step/.style={
    flow step,
    text = black,
    fill = data,
  }
}

\begin{tikzpicture}%
  \node[sw] (sw1) {};%
  \node[kernel] (k1) [right = of sw1.center] {};%
  \node[app] (a1) [left = of sw1.center] {};%
  \node[kumem] (amem1) at ($(a1.east) + (-0.8cm,-0.1cm)$) {};%

  \node[nic, label={[anchor = east]right:{\small NIC}}] (n1) [below = of a1.south west, anchor=west] {};
  \node[sw, node distance=(\basewidth + 0.6cm)] (sw2) [right = of k1] {};
  \coordinate (i1) at ($(amem1 |- n1.south) -(0,0.3cm) $);

  \node[control point, node distance = (\baseskip + 0.2cm)] (c1) [above right = of amem1] {};

  \node[control point] (c2n) at (n1-|c1) {};
  \draw[flow] (amem1) to node[data step] {4} (i1);
  \draw[control] (c1) --   node[control step] {3} (c2n);

\end{tikzpicture}%
        \caption{RDMA}%
        \label{fig:network-rdma}
    \end{subfigure}
    ~
    \begin{subfigure}[t]{0.3\linewidth}
        \tikzset{
sw/.style = {
    node distance = 0pt,
  },
user/.style = {
    top color=ulevel!60,
    bottom color=ulevel,
    rectangle,shade,
    rounded corners,
  },
kernel/.style = {
user,
minimum width = (\basewidth),
minimum height = \swheight,
node distance = \baseskip,
label={[anchor=base]above,label distance=0.7ex:{\emph{\small Kernel}}},
},
app/.style = {
user,
minimum width = (\basewidth),
minimum height = \swheight,
node distance = \baseskip,
label={[anchor=base]above,label distance=0.7ex:{\emph{\small User}}},
},
mem/.style = {
    minimum height=10pt,
    minimum width=10pt,
    node distance = 0pt,
    rounded corners=1pt,
  },
kmem/.style = {
    mem,
    fill=nicmem,
  },
kumem/.style = {
    umem,
    fill=nicmem,
  },
nic/.style = {
    user,
    top color=dlevel!40,
    bottom color=dlevel,
    node distance = 12pt,
    minimum width = (\basewidth + \basewidth + \baseskip + \baseskip),
    minimum height = 2.5ex
  },
flow/.style = {
rounded corners, line width = 1pt, color=data, -{Latex}
},
control/.style = {
rounded corners, line width = 0.8pt, color=control, -{Latex}
},
control point/.style = {
    draw,
    minimum size=3pt,
    align=center,
    inner sep=0pt,
    color=black,
    fill=black,
    node distance = 4pt,
    circle,
  },
flow step/.style= {
    draw,
    circle ,
    text=black,
    fill=control,
    inner sep=0.1pt,
    font=\footnotesize,
  },
control step/.style={
    flow step,
    text=black,
    fill=control,
  },
data step/.style={
    flow step,
    text = black,
    fill = data,
  }
}

\begin{tikzpicture}
  \node[sw] (sw1) {};
  \node[kernel] (k1) [right = of sw1.center] {};
  \node[app] (a1) [left = of sw1.center] {};

  \node[kumem] (amem1) at ($(a1.east) + (-0.3cm,-0.1cm)$) {};
  \coordinate (kmem1) at ($(k1.west) + (.5cm,-0.1cm)$) {};

  \node[nic, label={[anchor = west]left:{\small NIC}}] (n1) [below = of a1.south west, anchor=west] {};
  \node[sw, node distance=(\basewidth + 0.6cm)] (sw2) [right = of k1] {};
  \coordinate (i1) at ($(amem1 |- n1.south) -(0,0.3cm) $);

  \node[control point] (c1) [above left = of amem1] {};

  \coordinate (c2) at (k1|-c1) {};
  \node[control point] (c2n) at (n1-|c2) {};
  \coordinate (c3) at ($(c2 |- i1) -  (0,0.2cm)$);

  \coordinate (c4) at (c3|-i1);

  \draw[flow] (amem1) to node[data step] {4} (i1);

  \draw[control] (c1) --  (c2) --   node[rounded corners, control step] {3} (c2n);


\end{tikzpicture}%
        \caption{CoRD}%
        \label{fig:network-cord}
    \end{subfigure}
    \caption[Comparison of network architectures]{
        Socket, RDMA, and CoRD control- and data-planes.
        To send a message over a socket, the application invokes the high-level
        kernel network stack (\pict{control step}{1}).
        The kernel (\cref{fig:network-traditional}) or the application
        (\cref*{fig:network-rdma,fig:network-cord}) request the NIC driver to
        send the packet (\pict{control step}{3}), which copies the data from
        pinned (\pict{kmem}{}) memory to send it over the network (\pict{data
        step}{4}).
     }
    \label{fig:network-compare}
\end{figure}
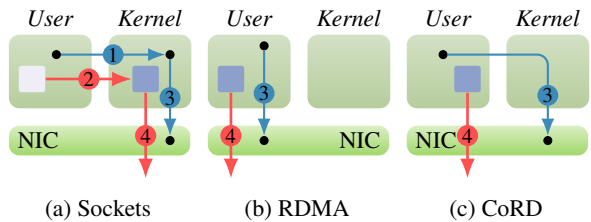

Our proposed architecture, CoRD integrates OS-level control into
high-performance user-level communication dataplane.
This section first provides an overview of both socket- and RDMA-based
communication (\cref{fig:network-compare}).
Then, it explains how CoRD modifies the RDMA dataplane to enable control over
the RDMA connections.

In a traditional socket-based API, as shown in \cref{fig:network-traditional},
communication starts with the application making a system call
(\pict{control step}{1}, e.g., \texttt{send}) which instructs the kernel to
dispatch a message over an existing connection (e.g., a TCP socket).
Subsequently, the kernel copies (\pict{data step}{2}) the message content into
NIC-accessible (\emph{pinned}, \pict{kmem}{}) memory.
Finally, the kernel \emph{triggers} the NIC (\pict{control step}{3}) to start
the transmission of message packets (\pict{data step}{4}).
This entire operation is managed by a complex kernel-level network stack,
designed to maintain scalability and improve resource utilization across a large
number of user applications.
However, this design philosophy inadvertently hampers network performance.

In contrast to socket-based communication, RDMA networks
(\cref{fig:network-rdma}) delegate a significant part of control to the
application, allowing it to manage pinned memory (\pict{kumem}{}) and access the
NIC directly (\pict{control step}{3}).
The NIC can retrieve the message content directly from the user memory
(\pict{data step}{4}), thus bypassing the kernel.
Note, that while this requires some network stack functionality at the
application level, a majority of responsibilities, such as managing concurrent
users, are offloaded to the NIC.
Despite the complexity added to the application, this architecture ultimately
yields the lowest possible communication latency.

CoRD principally mimics the classical RDMA architecture, retaining a majority of
its performance attributes.
The RDMA application keeps its responsibilities under CoRD, including the
management of pinned memory.
The drivers at the user level in RDMA (\cref{fig:network-rdma}) and at the
kernel level in CoRD (\cref{fig:network-cord}) are largely equivalent, thereby
ensuring a lightweight and transparently interchangeable layer between the
application and the NIC.
The critical distinction, however, lies in the NIC access.
CoRD mandates a system call from the application (\pict{control step}{3}),
thereby preventing it from directly accessing the NIC.

In CoRD each dataplane operation goes through the kernel, but the kernel
processing is kept to a minimum.
In contrast to socket-based communication, CoRD only permits the enforcement of
lightweight, non-blocking policies to maintain its high network performance.
However, this limitation is not significant, as without system-wide resource
sharing CoRD can afford to uphold simpler, more streamlined policies, which we
refer to as \emph{CoRD policies}.
CoRD policies should be powerful enough to implement QoS, security, and
isolation similarly to other dataplane interception techniques~\cite{freeflow,
masq}.
CoRD inevitably adds a constant per-message latency from user-kernel switching.
The overhead from the enforcement of CoRD policies depends greatly on the
specifics of the implemented functionality, which are beyond the scope of this
paper.

\section{Implementation}
\label{sec:implementation}

The goal of our prototype implementation is to measure the cost of CoRD's
architecture, which we expect to be higher than the bare minimum overhead of
user-kernel transition (see~\cref{sec:motivation}).
We modify the user- and kernel-level \texttt{mlx5} device drivers for NVIDIA's
ConnectX-series NICs.
These drivers are used by the \emph{ibverbs} library~\cite{ibverbs} to funnel
data-plane operations through the kernel.
The ibverbs API is the \enquote{narrow waist} of many high-performance
user-level network stacks~\cite{freeflow}, which use it either
directly~\cite{mitchellUsingOneSidedRDMA2013, farm, fasst} or through a
higher-level API~\cite{charm, legion, mpi-3}.
As this is a preliminary study, we do not attempt any performance optimizations,
so the resulting overhead represents the worst case.

The ibverbs API defines \emph{control- and data-plane} operations:
Control-plane operations set up communication; they register device-accessible
memory, create communication endpoints (\emph{queue pairs}), etc.
Data-plane operations initiate message sending (\lstinline{ibv_post_send}) and
receiving (\lstinline{ibv_post_recv}) or perform a non-blocking check if any of
these operations have completed (\lstinline{ibv_poll_cq}).
Control-plane operations require kernel support, whereas dataplane operations
usually bypass the kernel.

Applications request ibverbs control-plane operations using \lstinline{ioctl}
system call.
Unfortunately, the arguments to ibverbs calls are complex data structures that
must be \emph{serialized} and \emph{deserialized} to go through the
\lstinline|ioctl| interface.
These operations add to the system-call overhead but are not a performance
critical for control-plane operations.
%
On a positive side, the ioctl code path allows the OS to intercept control-plane
ibverbs calls and thus enforce security~\cite{selinux, apparmor},
isolation~\cite{paravpanditManpageRdmasystemRDMA2020}, and resource management
policies~\cite{cgroup}.
CoRD extends this existing functionality to data-plane operations.

After entering the kernel, the data-plane operation reaches the kernel-level
device driver, which normally provide high-performance networking
inside the Linux kernel~\cite{iser, lustre}.
We change the driver minimally, so it can work with the ibverbs objects created
by the user application.
Overall we added or modified \textasciitilde 250 lines in the kernel-level driver
and \textasciitilde 20 lines in the user-level driver.
Except for the kernel interposition, the ibverbs API remains unchanged and
introduces no interrupts or asynchronous invocations on the data plane.
In other words, without CoRD policies, the only overhead comes from crossing the
user-kernel boundary.

Our implementation depends on the NIC being able to access an application's
virtual memory as applications pass message buffers to the NIC by virtual
address.
This is a common feature for high-performance NICs~\cite{dsa, ibverbs}, which,
in contrast to other approaches (e.g., \lstinline{io_uring}~\cite{iouring}),
relieves the kernel from virtual-to-device address translations on the critical
path.
If the application passes an invalid address, the NIC returns an error but does
not access any memory that was not explicitly provided to the application.

\section{Evaluation}
\label{sec:eval}

Our goal is to evaluate how CoRD impacts raw communication performance as well
as end-to-end application performance.
We evaluate two systems.
First, a system \emph{L}, comprising two nodes, each with Intel i5-4590 4-core
CPUs and \SI{200}{\giga\bit/s} NVIDIA ConnectX-6 Dx RoCE NICs communicating at
up to \SI{100}{\giga\bit/s} due to motherboard limitations.
The nodes are back-to-back connected.
The system runs vanilla Linux 6.0.0-rc7 with or without our patch to support
CoRD in the \texttt{mlx5} driver.
We also disable Turbo Boost, pin all the benchmark processes to dedicated cores,
and set the CPU power governor to the highest performance mode.

Second, is a remote two-node system \emph{A} deployed in the Azure Cloud.
We use virtualized HB120 instances with two 64-core AMD EPYC 7V73X CPUs (only
120 cores passed to VM) and virtualized \SI{200}{\giga\bit/s} NVIDIA ConnectX-6
Infiniband NICs.
In both systems we disable KPTI~\cite{grussKASLRDeadLong2017, kpti}, an
expensive kernel-level Meltdown attack~\cite{meltdown} mitigation, because
modern CPUs do not need it.
Specifically, system A's CPU mitigates Meltdown in hardware.
We run the benchmarks in the same way as on system L, except for not being able
to disable dynamic frequency scaling due to the cloud provider policy.
%

\begin{figure}
    \centering
    \input{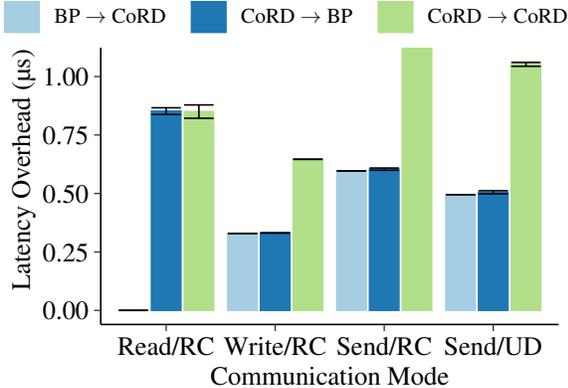} 
    \caption[Latency overhead at system L]%
    {%
        Latency overhead on system L (baseline in \cref{tab:remove-lat}) when
        communicating over different transports (RC/UD) using one-sided
        (Read/Write) or two-sided (Send) communication.
        Client and server can independently run bypass~(BP) or
        CoRD~(CD), \enquote{$\rightarrow$} indicates the direction
        of communication (from client to server).
    }%
    \label{fig:lat-overhead}
\end{figure}

We first measured the overhead CoRD adds to the point-to-point message latency,
using the perftest 4.5 benchmark suite~\cite{perftest}.
The communication can go either through Reliable Connection~(RC) or Unreliable
Datagram~(UD) transports.
In addition to two-sided send/receive communication, RC supports one-sided RDMA
read/write operations.
CoRD can independently be enabled on the sender or receiver, allowing us to
study which side contributes to latency more.
\Cref{fig:lat-overhead} shows the absolute latency overhead compared to
bypass-to-bypass communication, measured with a message size of
\SI{4}{\kibi\byte}.
We observed the same numbers for other message sizes.

When CoRD runs only at the server side, RDMA read has no overhead, because the
sender's CPU does not participate in sending the message.
In contrast, with RDMA write operations, perftest uses two writes to
exchange data:
One from the client to the server for synchronization, another for the server to
fetch the data from the client.
For operations except RDMA read, enabling CoRD contributes equally towards
overall overhead from each side.

\begin{figure}
    \centering
    \vspace{-3.8mm}
    \input{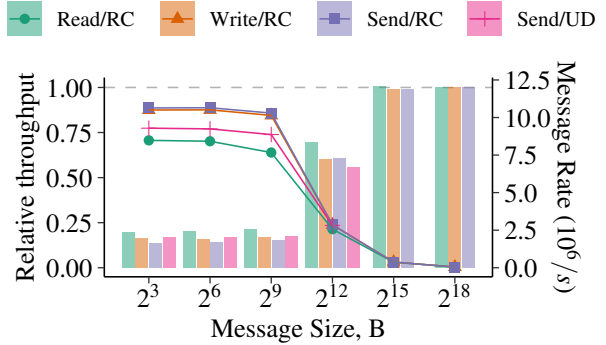}
    \vspace{-4mm}
    \caption[Relative throughput on system L]%
    {%
        CoRD's throughput on system L relative to bypass communication.
        Communication goes over RC or UD using one-sided (Read/Write) or
        two-sided (Send) operations.
        UD supports only up to \SI{4}{\kibi\byte} messages.
        The overlayed lines show the message rate (right axis) in the bypass configuration.
    }%
    \label{fig:bw-overhead}
\end{figure}

Constant per-message overhead results in significantly lower maximum throughput
for large bursts of small messages.
\Cref{fig:bw-overhead} corroborates this statement because, with larger
messages, bandwidth degradation becomes insignificant.
This behavior is similar for all types of communication (RC/UD,
Send/Read/Write) as the per-message overhead is similar.
Specifically, for \SI{32}{\kibi\byte} messages exchanged using send operations,
perftest measured \textasciitilde 370k messages per second and only $1\%$
bandwidth degradation. 
%

\begin{figure}[t]
    \centering
    \scalebox{.9}{
\begin{tikzpicture}[x=1pt,y=1pt]
\definecolor{fillColor}{RGB}{255,255,255}
\path[use as bounding box,fill=fillColor,fill opacity=0.00] (0,0) rectangle (238.49, 18.07);
\begin{scope}
\path[clip] (  0.00,  0.00) rectangle (238.49, 18.07);
\definecolor{fillColor}{RGB}{255,255,255}

\path[fill=fillColor] ( -5.87, -4.64) rectangle (244.36, 22.71);
\end{scope}
\begin{scope}
\path[clip] (  0.00,  0.00) rectangle (238.49, 18.07);
\definecolor{fillColor}{RGB}{27,158,119}

\path[fill=fillColor,fill opacity=0.50] (  4.84,  1.07) rectangle ( 20.76, 16.99);
\end{scope}
\begin{scope}
\path[clip] (  0.00,  0.00) rectangle (238.49, 18.07);
\definecolor{fillColor}{RGB}{217,95,2}

\path[fill=fillColor,fill opacity=0.50] ( 65.20,  1.07) rectangle ( 81.13, 16.99);
\end{scope}
\begin{scope}
\path[clip] (  0.00,  0.00) rectangle (238.49, 18.07);
\definecolor{fillColor}{RGB}{117,112,179}

\path[fill=fillColor,fill opacity=0.50] (125.23,  1.07) rectangle (141.15, 16.99);
\end{scope}
\begin{scope}
\path[clip] (  0.00,  0.00) rectangle (238.49, 18.07);
\definecolor{fillColor}{RGB}{231,41,138}

\path[fill=fillColor,fill opacity=0.50] (185.15,  1.07) rectangle (201.07, 16.99);
\end{scope}
\begin{scope}
\path[clip] (  0.00,  0.00) rectangle (238.49, 18.07);
\definecolor{drawColor}{RGB}{0,0,0}

\node[text=drawColor,anchor=base west,inner sep=0pt, outer sep=0pt, scale=  0.80] at ( 26.47,  6.28) {Read/RC};
\end{scope}
\begin{scope}
\path[clip] (  0.00,  0.00) rectangle (238.49, 18.07);
\definecolor{drawColor}{RGB}{0,0,0}

\node[text=drawColor,anchor=base west,inner sep=0pt, outer sep=0pt, scale=  0.80] at ( 86.84,  6.28) {Write/RC};
\end{scope}
\begin{scope}
\path[clip] (  0.00,  0.00) rectangle (238.49, 18.07);
\definecolor{drawColor}{RGB}{0,0,0}

\node[text=drawColor,anchor=base west,inner sep=0pt, outer sep=0pt, scale=  0.80] at (146.86,  6.28) {Send/RC};
\end{scope}
\begin{scope}
\path[clip] (  0.00,  0.00) rectangle (238.49, 18.07);
\definecolor{drawColor}{RGB}{0,0,0}

\node[text=drawColor,anchor=base west,inner sep=0pt, outer sep=0pt, scale=  0.80] at (206.79,  6.28) {Send/UD};
\end{scope}
\end{tikzpicture}}
    \begin{subcaptionblock}{.49\linewidth}
        \centering
\begin{tikzpicture}[x=1pt,y=1pt]
\definecolor{fillColor}{RGB}{255,255,255}
\path[use as bounding box,fill=fillColor,fill opacity=0.00] (0,0) rectangle (115.63,130.09);
\begin{scope}
\path[clip] (  0.00,  0.00) rectangle (115.63,130.09);
\definecolor{drawColor}{RGB}{255,255,255}
\definecolor{fillColor}{RGB}{255,255,255}

\path[draw=drawColor,line width= 0.5pt,line join=round,line cap=round,fill=fillColor] (  0.00,  0.00) rectangle (115.63,130.09);
\end{scope}
\begin{scope}
\path[clip] ( 33.61, 29.66) rectangle (110.63,125.09);
\definecolor{fillColor}{RGB}{255,255,255}

\path[fill=fillColor] ( 33.61, 29.66) rectangle (110.63,125.09);
\definecolor{fillColor}{RGB}{27,158,119}

\path[fill=fillColor,fill opacity=0.50] ( 37.11, 34.00) rectangle ( 44.66, 89.88);
\definecolor{fillColor}{RGB}{217,95,2}

\path[fill=fillColor,fill opacity=0.50] ( 45.50, 34.00) rectangle ( 53.05, 86.22);
\definecolor{fillColor}{RGB}{117,112,179}

\path[fill=fillColor,fill opacity=0.50] ( 53.89, 34.00) rectangle ( 61.44,114.86);
\definecolor{fillColor}{RGB}{231,41,138}

\path[fill=fillColor,fill opacity=0.50] ( 62.28, 34.00) rectangle ( 69.83,106.21);
\definecolor{fillColor}{RGB}{27,158,119}

\path[fill=fillColor,fill opacity=0.50] ( 74.40, 34.00) rectangle ( 81.96, 86.29);
\definecolor{fillColor}{RGB}{217,95,2}

\path[fill=fillColor,fill opacity=0.50] ( 82.79, 34.00) rectangle ( 90.35, 71.02);
\definecolor{fillColor}{RGB}{117,112,179}

\path[fill=fillColor,fill opacity=0.50] ( 91.19, 34.00) rectangle ( 98.74, 99.62);
\definecolor{fillColor}{RGB}{231,41,138}

\path[fill=fillColor,fill opacity=0.50] ( 99.58, 34.00) rectangle (107.13, 91.20);
\definecolor{drawColor}{RGB}{0,0,0}

\path[draw=drawColor,line width= 0.6pt,line join=round] ( 37.11, 91.01) --
	( 44.66, 91.01);

\path[draw=drawColor,line width= 0.6pt,line join=round] ( 40.88, 91.01) --
	( 40.88, 88.75);

\path[draw=drawColor,line width= 0.6pt,line join=round] ( 37.11, 88.75) --
	( 44.66, 88.75);

\path[draw=drawColor,line width= 0.6pt,line join=round] ( 45.50, 93.10) --
	( 53.05, 93.10);

\path[draw=drawColor,line width= 0.6pt,line join=round] ( 49.27, 93.10) --
	( 49.27, 79.33);

\path[draw=drawColor,line width= 0.6pt,line join=round] ( 45.50, 79.33) --
	( 53.05, 79.33);

\path[draw=drawColor,line width= 0.6pt,line join=round] ( 53.89,120.75) --
	( 61.44,120.75);

\path[draw=drawColor,line width= 0.6pt,line join=round] ( 57.67,120.75) --
	( 57.67,108.97);

\path[draw=drawColor,line width= 0.6pt,line join=round] ( 53.89,108.97) --
	( 61.44,108.97);

\path[draw=drawColor,line width= 0.6pt,line join=round] ( 62.28,112.42) --
	( 69.83,112.42);

\path[draw=drawColor,line width= 0.6pt,line join=round] ( 66.06,112.42) --
	( 66.06, 99.99);

\path[draw=drawColor,line width= 0.6pt,line join=round] ( 62.28, 99.99) --
	( 69.83, 99.99);

\path[draw=drawColor,line width= 0.6pt,line join=round] ( 74.40, 89.33) --
	( 81.96, 89.33);

\path[draw=drawColor,line width= 0.6pt,line join=round] ( 78.18, 89.33) --
	( 78.18, 83.25);

\path[draw=drawColor,line width= 0.6pt,line join=round] ( 74.40, 83.25) --
	( 81.96, 83.25);

\path[draw=drawColor,line width= 0.6pt,line join=round] ( 82.79, 77.34) --
	( 90.35, 77.34);

\path[draw=drawColor,line width= 0.6pt,line join=round] ( 86.57, 77.34) --
	( 86.57, 64.70);

\path[draw=drawColor,line width= 0.6pt,line join=round] ( 82.79, 64.70) --
	( 90.35, 64.70);

\path[draw=drawColor,line width= 0.6pt,line join=round] ( 91.19,105.60) --
	( 98.74,105.60);

\path[draw=drawColor,line width= 0.6pt,line join=round] ( 94.96,105.60) --
	( 94.96, 93.64);

\path[draw=drawColor,line width= 0.6pt,line join=round] ( 91.19, 93.64) --
	( 98.74, 93.64);

\path[draw=drawColor,line width= 0.6pt,line join=round] ( 99.58, 96.42) --
	(107.13, 96.42);

\path[draw=drawColor,line width= 0.6pt,line join=round] (103.35, 96.42) --
	(103.35, 85.99);

\path[draw=drawColor,line width= 0.6pt,line join=round] ( 99.58, 85.99) --
	(107.13, 85.99);
\end{scope}
\begin{scope}
\path[clip] (  0.00,  0.00) rectangle (115.63,130.09);
\definecolor{drawColor}{RGB}{0,0,0}

\path[draw=drawColor,line width= 0.6pt,line join=round] ( 33.61, 29.66) --
	( 33.61,125.09);
\end{scope}
\begin{scope}
\path[clip] (  0.00,  0.00) rectangle (115.63,130.09);
\definecolor{drawColor}{RGB}{0,0,0}

\node[text=drawColor,anchor=base east,inner sep=0pt, outer sep=0pt, scale=  1.00] at ( 29.11, 30.56) {0.0};

\node[text=drawColor,anchor=base east,inner sep=0pt, outer sep=0pt, scale=  1.00] at ( 29.11, 50.04) {0.5};

\node[text=drawColor,anchor=base east,inner sep=0pt, outer sep=0pt, scale=  1.00] at ( 29.11, 69.52) {1.0};

\node[text=drawColor,anchor=base east,inner sep=0pt, outer sep=0pt, scale=  1.00] at ( 29.11, 89.01) {1.5};

\node[text=drawColor,anchor=base east,inner sep=0pt, outer sep=0pt, scale=  1.00] at ( 29.11,108.49) {2.0};
\end{scope}
\begin{scope}
\path[clip] (  0.00,  0.00) rectangle (115.63,130.09);
\definecolor{drawColor}{gray}{0.20}

\path[draw=drawColor,line width= 0.5pt,line join=round] ( 31.11, 34.00) --
	( 33.61, 34.00);

\path[draw=drawColor,line width= 0.5pt,line join=round] ( 31.11, 53.48) --
	( 33.61, 53.48);

\path[draw=drawColor,line width= 0.5pt,line join=round] ( 31.11, 72.97) --
	( 33.61, 72.97);

\path[draw=drawColor,line width= 0.5pt,line join=round] ( 31.11, 92.45) --
	( 33.61, 92.45);

\path[draw=drawColor,line width= 0.5pt,line join=round] ( 31.11,111.93) --
	( 33.61,111.93);
\end{scope}
\begin{scope}
\path[clip] (  0.00,  0.00) rectangle (115.63,130.09);
\definecolor{drawColor}{RGB}{0,0,0}

\path[draw=drawColor,line width= 0.6pt,line join=round] ( 33.61, 29.66) --
	(110.63, 29.66);
\end{scope}
\begin{scope}
\path[clip] (  0.00,  0.00) rectangle (115.63,130.09);
\definecolor{drawColor}{gray}{0.20}

\path[draw=drawColor,line width= 0.5pt,line join=round] ( 53.47, 27.16) --
	( 53.47, 29.66);

\path[draw=drawColor,line width= 0.5pt,line join=round] ( 90.77, 27.16) --
	( 90.77, 29.66);
\end{scope}
\begin{scope}
\path[clip] (  0.00,  0.00) rectangle (115.63,130.09);
\definecolor{drawColor}{RGB}{0,0,0}

\node[text=drawColor,anchor=base,inner sep=0pt, outer sep=0pt, scale=  1.00] at ( 53.47, 18.28) {$2^{6}$};

\node[text=drawColor,anchor=base,inner sep=0pt, outer sep=0pt, scale=  1.00] at ( 90.77, 18.28) {$2^{12}$};
\end{scope}
\begin{scope}
\path[clip] (  0.00,  0.00) rectangle (115.63,130.09);
\definecolor{drawColor}{RGB}{0,0,0}

\node[text=drawColor,anchor=base,inner sep=0pt, outer sep=0pt, scale=  1.00] at ( 72.12,  6.94) {Message size, B};
\end{scope}
\begin{scope}
\path[clip] (  0.00,  0.00) rectangle (115.63,130.09);
\definecolor{drawColor}{RGB}{0,0,0}

\node[text=drawColor,rotate= 90.00,anchor=base,inner sep=0pt, outer sep=0pt, scale=  1.00] at ( 11.89, 77.37) {Latency overhead (µs)};
\end{scope}
\end{tikzpicture} 
        \caption{Latency overhead}\label{fig:azure-lat}
    \end{subcaptionblock}%
    \begin{subcaptionblock}{.49\linewidth}
        \centering
\begin{tikzpicture}[x=1pt,y=1pt]
\definecolor{fillColor}{RGB}{255,255,255}
\path[use as bounding box,fill=fillColor,fill opacity=0.00] (0,0) rectangle (115.63,130.09);
\begin{scope}
\path[clip] (  0.00,  0.00) rectangle (115.63,130.09);
\definecolor{drawColor}{RGB}{255,255,255}
\definecolor{fillColor}{RGB}{255,255,255}

\path[draw=drawColor,line width= 0.5pt,line join=round,line cap=round,fill=fillColor] (  0.00,  0.00) rectangle (115.63,130.09);
\end{scope}
\begin{scope}
\path[clip] ( 33.61, 29.66) rectangle (110.63,125.09);
\definecolor{fillColor}{RGB}{255,255,255}

\path[fill=fillColor] ( 33.61, 29.66) rectangle (110.63,125.09);
\definecolor{drawColor}{RGB}{0,0,0}

\path[draw=drawColor,draw opacity=0.30,line width= 0.6pt,dash pattern=on 4pt off 4pt ,line join=round] ( 33.61,120.52) -- (110.63,120.52);
\definecolor{fillColor}{RGB}{27,158,119}

\path[fill=fillColor,fill opacity=0.50] ( 37.11, 34.00) rectangle ( 42.24, 66.63);
\definecolor{fillColor}{RGB}{217,95,2}

\path[fill=fillColor,fill opacity=0.50] ( 42.81, 34.00) rectangle ( 47.93, 58.28);
\definecolor{fillColor}{RGB}{117,112,179}

\path[fill=fillColor,fill opacity=0.50] ( 48.50, 34.00) rectangle ( 53.63, 58.50);
\definecolor{fillColor}{RGB}{231,41,138}

\path[fill=fillColor,fill opacity=0.50] ( 54.20, 34.00) rectangle ( 59.33, 55.08);
\definecolor{fillColor}{RGB}{27,158,119}

\path[fill=fillColor,fill opacity=0.50] ( 65.28, 34.00) rectangle ( 70.41, 99.77);
\definecolor{fillColor}{RGB}{217,95,2}

\path[fill=fillColor,fill opacity=0.50] ( 70.98, 34.00) rectangle ( 76.11, 94.57);
\definecolor{fillColor}{RGB}{117,112,179}

\path[fill=fillColor,fill opacity=0.50] ( 76.68, 34.00) rectangle ( 81.81, 91.91);
\definecolor{fillColor}{RGB}{27,158,119}

\path[fill=fillColor,fill opacity=0.50] ( 90.61, 34.00) rectangle ( 95.73,120.75);
\definecolor{fillColor}{RGB}{217,95,2}

\path[fill=fillColor,fill opacity=0.50] ( 96.30, 34.00) rectangle (101.43,120.52);
\definecolor{fillColor}{RGB}{117,112,179}

\path[fill=fillColor,fill opacity=0.50] (102.00, 34.00) rectangle (107.13,120.42);
\end{scope}
\begin{scope}
\path[clip] (  0.00,  0.00) rectangle (115.63,130.09);
\definecolor{drawColor}{RGB}{0,0,0}

\path[draw=drawColor,line width= 0.6pt,line join=round] ( 33.61, 29.66) --
	( 33.61,125.09);
\end{scope}
\begin{scope}
\path[clip] (  0.00,  0.00) rectangle (115.63,130.09);
\definecolor{drawColor}{RGB}{0,0,0}

\node[text=drawColor,anchor=base east,inner sep=0pt, outer sep=0pt, scale=  1.00] at ( 29.11, 30.56) {0.0};

\node[text=drawColor,anchor=base east,inner sep=0pt, outer sep=0pt, scale=  1.00] at ( 29.11, 73.81) {0.5};

\node[text=drawColor,anchor=base east,inner sep=0pt, outer sep=0pt, scale=  1.00] at ( 29.11,117.07) {1.0};
\end{scope}
\begin{scope}
\path[clip] (  0.00,  0.00) rectangle (115.63,130.09);
\definecolor{drawColor}{gray}{0.20}

\path[draw=drawColor,line width= 0.5pt,line join=round] ( 31.11, 34.00) --
	( 33.61, 34.00);

\path[draw=drawColor,line width= 0.5pt,line join=round] ( 31.11, 77.26) --
	( 33.61, 77.26);

\path[draw=drawColor,line width= 0.5pt,line join=round] ( 31.11,120.52) --
	( 33.61,120.52);
\end{scope}
\begin{scope}
\path[clip] (  0.00,  0.00) rectangle (115.63,130.09);
\definecolor{drawColor}{RGB}{0,0,0}

\path[draw=drawColor,line width= 0.6pt,line join=round] ( 33.61, 29.66) --
	(110.63, 29.66);
\end{scope}
\begin{scope}
\path[clip] (  0.00,  0.00) rectangle (115.63,130.09);
\definecolor{drawColor}{gray}{0.20}

\path[draw=drawColor,line width= 0.5pt,line join=round] ( 48.22, 27.16) --
	( 48.22, 29.66);

\path[draw=drawColor,line width= 0.5pt,line join=round] ( 73.54, 27.16) --
	( 73.54, 29.66);

\path[draw=drawColor,line width= 0.5pt,line join=round] ( 98.87, 27.16) --
	( 98.87, 29.66);
\end{scope}
\begin{scope}
\path[clip] (  0.00,  0.00) rectangle (115.63,130.09);
\definecolor{drawColor}{RGB}{0,0,0}

\node[text=drawColor,anchor=base,inner sep=0pt, outer sep=0pt, scale=  1.00] at ( 48.22, 18.28) {$2^{12}$};

\node[text=drawColor,anchor=base,inner sep=0pt, outer sep=0pt, scale=  1.00] at ( 73.54, 18.28) {$2^{14}$};

\node[text=drawColor,anchor=base,inner sep=0pt, outer sep=0pt, scale=  1.00] at ( 98.87, 18.28) {$2^{16}$};
\end{scope}
\begin{scope}
\path[clip] (  0.00,  0.00) rectangle (115.63,130.09);
\definecolor{drawColor}{RGB}{0,0,0}

\node[text=drawColor,anchor=base,inner sep=0pt, outer sep=0pt, scale=  1.00] at ( 72.12,  6.94) {Message size};
\end{scope}
\begin{scope}
\path[clip] (  0.00,  0.00) rectangle (115.63,130.09);
\definecolor{drawColor}{RGB}{0,0,0}

\node[text=drawColor,rotate= 90.00,anchor=base,inner sep=0pt, outer sep=0pt, scale=  1.00] at ( 11.89, 77.37) {Relative throughput};
\end{scope}
\end{tikzpicture} 
        \caption{Throughput}\label{tab:azure-bw}
    \end{subcaptionblock}%
    \caption[Latency and throughput on system A]%
    {%
        Latency overhead and relative throughput on system A when
        communicating over different transports (RC/UD) using
        one-sided (Read/Write) or two-sided (Send) communication.
    }%
    \label{fig:azure-overhead}
\end{figure}

\begin{figure}
    \centering
    \input{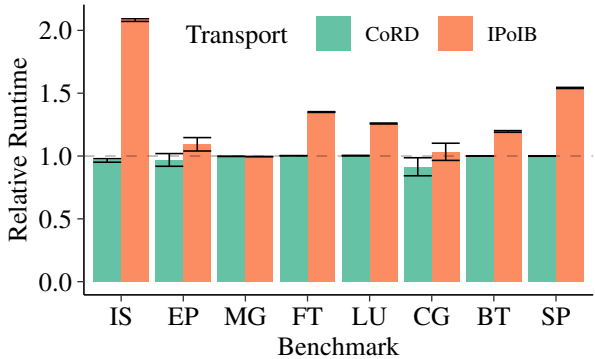}
    \caption[Bandwidth overhead at system A]%
    {%
        Relative runtime of the NPB benchmarks on system A.
    }%
    \label{fig:npb-overhead}
\end{figure}

System A exhibits similar behavior (\cref{fig:azure-overhead}), except for a few
minor differences:
Overall per-message overhead is larger, has higher variation, and has two
statistical modes (\cref{fig:azure-lat} shows them) for small ($\leq
\SI{1}{\kibi\byte}$) and large messages, with the larger messages having smaller
overhead.
%
%
%
%
Our initial investigation suggests that CoRD has higher overhead for small
messages, because current implementation of CoRD lacks support for inline
messages, whereas baseline supports them.
At the same message size, system L shows a higher throughput reduction than system
A because its network has double the bandwidth.
Nevertheless, bandwidth reduction becomes negligible from a certain message size
(see~\cref{tab:azure-bw}).

To estimate the effect on real-world applications, we measure the performance of
the MPI version of the NPB suite~\cite{npb}.
We compare communication over RDMA, CoRD, and IPoIB.
To amplify the network effects, we bar the MPI library from using shared memory
communication.
We picked IPoIB for comparison because it communicates over Infiniband NIC, yet
offers fine-grained control over data-plane operations, making it a functionally
equivalent competitor to CoRD.
Each benchmark has limitations on the number of processes allowed for a run,
which in our case ranged from \numrange{128}{240}.

For all the benchmarks, CoRD has nearly zero overhead over baseline
kernel-bypass communication whereas IPoIB is up to $2\times$ slower.
IPoIB is the slowest with the IS (integer sorting) and SP (matrix factorization)
benchmarks, which are simultaneously data intensive (each process sends
\SI{72}{\giga\bit/\second} and \SI{34}{\giga\bit/\second}, respectively) and
message intensive ($\approx$\SI{1300}{messages/second} per process).
EP (embarrassingly parallel), which communicates very little, and CG (conjugate
gradient), which communicates using few large messages, see a slight performance
boost with CoRD.
Similarly to CG, we observe CoRD marginally outperforming kernel bypass in
large-message bandwidth microbenchmarks when Turbo Boost is enabled.
This behavior suggests that system calls interact with DVFS.
Overall, our initial implementation of CoRD has negligible overhead, even for
large-scale real-world applications.


\section{Discussion and Conclusion}

Modern cloud systems predominantly rely on socket-based networking to manage a
myriad of distributed applications.
Key to this operation is the precise control over application communication
channels using tools such as Linux packet filtering~\cite{weave} or
eBPF~\cite{calico-ebpf}.
Despite persistent enhancements in high-performance TCP/IP~\cite{netchannel,
tsor}, it still falls significantly short of the performance of RDMA networks.

High-performance networking is diverse and complex~\cite{dpdk, ugni, ibverbs,
barrettPortalsNetworkProgramming2017, snap}, so, unsurprisingly, there are no
\enquote{one size fits all}
solutions~\cite{taranovEfficientRDMACommunication2022}.
Striving for optimal problem-specific solutions, researchers abandoned
polling~\cite{venkateshCaseApplicationobliviousEnergyefficient2015,
bierbaumEfficientOversubscriptionCost2022}, zero-copy~\cite{mvapich2,
taranovEfficientRDMACommunication2022}, lossless congestion
control~\cite{liranlissLinuxSoftRoCEDriver2017, iwarp}, hardware
offloading~\cite{TAS, xdp, toe, arrakis}, and one-sided
operations~\cite{suRFPWhenRPC2017,kaliaUsingRDMAEfficiently2014,
dragojevicRDMAReadsUse2017} to achieve higher flexibility and resource
utilization.
So, giving up performance-enhancing features is not radically new.

Still, the high-performance networking community adamantly avoids
putting the OS kernel on the data path.
Instead, existing works add another device on the path of a packet, either in
the form of dedicated CPU cores~\cite{freeflow,iouring,
boyd-wickizerCoreyOperatingSystem2008, ousterhoutShenangoAchievingHigh2019,
tsaiLITEKernelRDMA2017, snap} or by offloading complicated data path
interception logic to the NIC.
Such offloading requires either an expensive SmartNIC~\cite{accelnet, spin,
sadokWeNeedKernel2021, fractos} or hardware modifications~\cite{ibverbs-odp,
lesokhinPageFaultSupport2017, migros}.
%
%
%
%

%
%
Our work is based on the idea that the actual problem with system calls is not
the cost of user-kernel transitions in hardware, but rather an obsolete and
inefficient API; in this case POSIX.
This idea is in line with the existing body of work~\cite{demikernel, fastcalls,
weiKRCOREMicrosecondscaleRDMA2021, pekkaenbergTranscendingPOSIXEnd2022,
atlidakisPOSIXAbstractionsModern2016}, and our early results support the
observation of kernel bypass being dispensable.
%
%
%

In the future, we strive for a smaller per-message overhead and incorporate
several policies, which we will evaluate with a modern bare-metal system.
We intend to assemble a set of real-world benchmark applications that shows the
\enquote{breaking point} of CoRD.
%
%
When successful, CoRD concepts can help in other domains like high-performance
storage~\cite{spdk,intelcorporationOneAPISpecification2rev1}, where APIs are
built on similar concepts.
%


\section*{Acknowledgments}

The research and the work presented in this paper has been supported by the
German priority program 1648 \enquote{Software for Exascale Computing} via the
research project FFMK~\cite{ffmk-web} (HA-2461/10-2).
The authors acknowledge the financial support by the Federal Ministry of
Education and Research of Germany in the programme of \enquote{Souverän.
Digital. Vernetzt.}.
Joint project 6G-life, project identification number: 16KISK001K
This work was supported in part by the German Research Foundation (DFG) within
the Collaborative Research Center HAEC and the the Center for Advancing
Electronics Dresden (cfaed).
The authors acknowledge support from the Oracle for Research for providing cloud
computing resources in project 172.
We would like to thank the anonymous HotOS and WORDS reviewers for their
valuable comments. 



\bibliographystyle{plain}
\bibliography{hotos2023}

\end{document}